\documentstyle[12pt]{article} 
\def\bbox{{\,\lower0.9pt\vbox{\hrule \hbox{\vrule height 0.2 cm

\hskip 0.2 cm

\vrule  height 0.2 cm}\hrule}\,}}

\def\bbox{{\,\lower0.9pt\vbox{\hrule \hbox{\vrule height 0.2 cm

\hskip 0.2 cm

\vrule  height 0.2 cm}\hrule}\,}}
\begin{document}
\setlength{\unitlength}{1mm}
\title{{\hfill {\small  } } \vspace*{2cm} \\
Statistical Mechanics
of Charged Black Holes in Induced Einstein-Maxwell
Gravity}
\author{\\
V.P. Frolov and D.V. Fursaev
\date{}}
\maketitle
\noindent  { { \em Theoretical Physics Institute,
Department of Physics, \ University of Alberta \\
Edmonton, Canada T6G
2J1 and } \\
{\em Joint Institute for Nuclear Research, Bogoliubov
Laboratory of Theoretical Physics, \\
141 980 Dubna, Russia} \\ \\
e-mail: frolov@phys.ualberta.ca, fursaev@thsun1.jinr.ru
}
\bigskip

\begin{abstract}
The statistical origin of the entropy of 
charged black holes in models of induced Einstein-Maxwell
gravity is investigated. The constituents inducing the Einstein-Maxwell 
action are charged and interact with an external gauge potential. 
This new feature, however, does not change
divergences of the statistical-mechanical entropy
of the constituents near the horizon. 
It is demonstrated that the mechanism of generation of the
Bekenstein-Hawking entropy in induced gravity 
is universal and it is basically the same
for charged and neutral black holes.
The concrete computations are 
carried out for induced Einstein-Maxwell gravity with a negative
cosmological constant in three space-time dimensions.
\end{abstract}

\bigskip

{\it PACS number(s): 04.60.+n, 12.25.+e, 97.60.Lf, 11.10.Gh}

\baselineskip=.6cm

\newpage

\section{Introduction}
\setcounter{equation}0

The Bekenstein-Hawking entropy $S^{BH}$ is one of the most
intriguing features of black holes. It is generally believed that
it is impossible 
to find its statistical-mechanical explanation 
in the framework of the classical Einstein gravity. It is
more likely that $S^{BH}$ hints to a 
more fundamental theory of quantum gravity which provides black holes
with microscopic degrees of freedom. 
Such a theory may be quite complicated, like the string
theory. Yet one may expect that
the mechanism of generation of $S^{BH}$ does not depend on the
details and is
universal.
One of the possibilities to understand this mechanism 
\cite{Jacobson}--\cite{FF:99a}
is to use the idea of Sakharov's induced gravity \cite{Sakharov}.
Sakharov's basic assumption is that the gravity becomes dynamical
as a result  of quantum effects of constituent fields.
In the models of induced gravity  
the Bekenstein-Hawking entropy $S^{BH}$ of a black hole has the 
following statistical-mechanical form 
\begin{equation}\label{i.1} 
S^{BH}=S^{SM}-Q~~~.
\end{equation}
Here $S^{SM}$ is the statistical-mechanical (entanglement)
entropy of the constituents located near the horizon and $Q$ is the 
Noether charge which appears because of non-minimal couplings.  
Relation (\ref{i.1}) has been demonstrated for static 
\cite{FFZ}--\cite{FF:98b} and 
rotating \cite{FF:99a} black holes in four space-time dimensions.  
These works however considered a pure induced gravity and, 
hence, the black hole solutions had no charges. In a 
more realistic situation one may expect that all other long-range 
fields are induced along with the gravitational one on an equal 
footing. By extending in this way Sakharov's assumption one could
model a fundamental theory which
unifies the gravity with other forces of the Nature and investigate
a more general class of black holes.  

The aim of the present paper  is to consider an induced
Einstein-Maxwell gravity, as a simplest model of such a unified 
theory.  
In this picture, the graviton $g_{\mu\nu}$ and photon $A_\mu$
are not fundamental, and are collective excitations of
the constituent fields.
As we will see, the constituents now are to be charged and 
interact with $A_\mu$ as with an external potential.
The action for $A_\mu$ 
is completely induced by the vacuum polarization 
and in the low-energy limit it coincides with the Maxwell action.  
In principle, 
the additional interaction of the constituents with the black 
hole electromagnetic field at the horizon can change the statistical 
entropy $S^{SM}$.  We show, however, that this interaction 
does not affect the divergent part of $S^{SM}$ and in the low-energy
limit relation (\ref{i.1}) for charged black holes preserve its form. 
Thus, the mechanism of generation
of the Bekenstein-Hawking entropy by the 
constituents seems to be universal
and not depending on the type of a black hole.

To simplify the analysis we 
study induced Einstein-Maxwell  gravity 
in three dimensions, more exactly, the Einstein-Maxwell gravity with
a negative cosmological constant.
The main interest to this theory is that it admits 
charged black holes \cite{BTZ},\cite{Clement}
and mimics some properties of the 
four-dimensional theory. There are also other reasons which motivate
our choice.
First, in three dimensions, one can easily construct 
induced gravity models
which are completely 
free from ultraviolet divergences while
in four-dimensions eliminating all the divergences becomes
a complicated problem. Second, a special class of
solutions in three-dimensional gravity are BTZ black holes \cite{BTZ}
whose entropy admits an alternative statistical representation in terms
of a conformal field theory \cite{Strominger}.

\bigskip

The paper is organized as follows.
In Section 2 we construct a model of induced
Einstein-Maxwell gravity with negative cosmological 
constant in three-dimensional
space-time. The thermodynamics of charged black holes
in this theory is discussed in Section 3.
In Section 4
we consider the properties charged fields near a 
charged black hole. We show 
that for non-extremal black holes
with a weak electric field the gauge interaction
near the horizon slightly shifts the mass of
a particle. This cause, however, a very little effect on
the density of energy levels of the field. 
By using this result in Section 5 we 
investigate the statistical-mechanical
origin of the Bekenstein-Hawking entropy 
of charged black holes in three-dimensional induced
Einstein-Maxwell gravity. 
Our concluding remarks in Section 6 concern charged black holes
in four-dimensional induced Einstein-Maxwell gravity.
Some details regarding computation of the induced action
can be found in Appendix.

\section{Induced Einstein-Maxwell gravity}
\setcounter{equation}0

The classical Einstein-Maxwell gravity is the theory of
the interacting
gravitational field, $g_{\mu\nu}$, and the Abelian gauge field
$A_\mu$. The corresponding (diffeo- and gauge-invariant) action is
\begin{equation}\label{4.1}
I[g,A]=\frac 14 \int d^3x\sqrt{-g}\left[{1 \over 4\pi G}(R-2\Lambda)
-F^{\mu\nu}F_{\mu\nu}\right]~~~.
\end{equation}
We consider this theory in three dimensions.
In (\ref{4.1}), $R$ is the scalar curvature defined for the
metric $g_{\mu\nu}$,
$F_{\mu\nu}=\nabla_\mu A_\nu-\nabla_\nu A_\mu$ is the Maxwell
strength tensor,
$G$ and $\Lambda$ are the gravitational (Newton) and
cosmological constants, respectively.
If the cosmological constant is negative, one of the
solutions of (\ref{4.1})  is anti-de Sitter space-time. 
For this reason and for the brevity we will call
such a theory AdSM-gravity (anti-de Sitter-Maxwell gravity).

\bigskip

In the induced gravity approach action (\ref{4.1}) is
generated as an effective action for a system of quantum
constituents.
Constituents of our model are charged massive scalar and spinor
fields 
on the space-time with the metric $g_{\mu\nu}$.
The fields do not interact to each other but interact to the
gauge field $A_\mu$. The theory is described by the quantum
action
\begin{equation}\label{4.2}
\Gamma[g,A]=\sum_s W_s[g,A]+\sum_d W_d[g,A]~~~,
\end{equation}
\begin{equation}\label{4.3}
W_s[g,A]=\frac i2 \log \det \left(-D_\mu D^\mu+\xi_s R+m_s^2\right)~~~,
\end{equation}
\begin{equation}\label{4.4}
W_d[g,A]=-i\log\det\left(\gamma^\mu D_{\mu}+m_d\right)~~~.
\end{equation}
At least some of scalars are non-minimally coupled
with the corresponding constants $\xi_s$. We have $N_s$,
scalars with masses $m_s$ and $N_d$ spinors with masses $m_d$.
The covariant derivative for the $k$-th constituent
with the charge $e_k$ is
\begin{equation}\label{4.5}
D_\mu=\nabla_\mu+e_k A_\mu~~~.
\end{equation}
Note that in three dimensions $A_\mu$  has the dimensionality of
$(\mbox{mass})^{1/2}$ and, thus, the elementary charges $e_k$
have a nontrivial dimensionality of
$(\mbox{mass})^{1/2}$.

The theory (\ref{4.2}) in three dimensions
has a very important property. It is completely free from the
ultraviolet divergences if the parameters of the model
subject to the constraints (see Appendix)
\begin{equation}\label{4.6}
2N_s-2N_d=0~~,~~2\sum_s m_s^2-2\sum_d m_d^2=0~~,~~
2N_s+N_d-12\sum_s \xi_s=0~~~.
\end{equation}
In three dimension the
spinors have two components and for this reason
equations (\ref{4.6}) coincide with the constraints
which provide finiteness of induced $2D$ gravity
\cite{FFGK}.

Suppose that all masses have the order of magnitude of
some specific scale $M$.
Now if the curvature of the space-time 
and the strength of the gauge field are small, i.e.,
\begin{equation}\label{4.7}
|R|\ll M^2~~~,~~~|F^{\mu\nu} F_{\mu\nu}|\ll M^3~~~,
\end{equation}
action (\ref{4.2}) can be approximated
by local decomposition in the curvature and the strength
tensor. In the leading approximation it coincides with
classical action (\ref{4.1})
\begin{equation}\label{4.8}
\Gamma[g,A]\simeq I[g,A]~~~,
\end{equation}
where the induced gravitational and cosmological
constants are determined by the parameters
of the quantum constituents (see Appendix for details)
\begin{equation}\label{4.9}
{1 \over G}=\frac 13 \left(2\sum_s(6\xi_s-1)m_s-\sum_d m_d\right)~~~,
\end{equation}
\begin{equation}\label{4.10}
{\Lambda \over G}={8 \over 3}\left(\sum_d m_d^3
-\sum_s m_s^3\right)~~~.
\end{equation}
It follows from (\ref{4.6}) and (\ref{4.10}) that the induced
gravity requires that at least some scalar constituents  are
non-minimally coupled with positive parameters $\xi_s$ in
order to provide positivity of $G$. Such models can be constructed
for the certain choices of $m_s$, $m_d$ and $\xi_s$. In what follows
we will also assume that the induced cosmological constant is negative.

\bigskip

It should be noted, that although the constraints
in 2D and 3D induced gravities are the same and the theories are
free from all the divergencies
there is an important difference between them.
The action of induced 2D gravity formulated
in form (\ref{4.1}) does not provide dynamical equations
for the metric. That is why one has to introduce
massless constituents to get the induced 2D gravity in the Liouville form
\cite{FFGK}. The attractive feature of 3D gravity (\ref{4.1}) is that
its properties are similar
to the Einstein-Maxwell gravity in four dimensions
and it has a number of interesting black hole
solutions.

\section{Thermodynamics of charged black holes in AdSM gravity}
\setcounter{equation}0

A static charged black hole solution in 3D AdSM-gravity (\ref{4.1})
was found in \cite{BTZ}. It can be written in the form
\begin{equation}\label{4.11}
ds^2=-N^2dt^2+N^{-2}dr^2+r^2d\varphi^2~~~,
\end{equation}
\begin{equation}\label{4.14}
N^2={1 \over a^2}(r^2-r_+^2) - {2G \over \pi} q^2\ln {r \over r_+}~~~,
\end{equation}
where $a^2=-1/\Lambda$.
The point $r=r_+$ corresponds to the black hole horizon, which
in three dimensions is a circle.
The corresponding gauge potential is
\begin{equation}\label{4.15}
A(r)=-{q \over 2\pi} \ln {r \over r_+}dt~~~,
\end{equation}
where $q$ is the charge of the black hole.
We impose condition $A_t(r_+)=0$ which provides regularity of
the vector field on the horizon.
One can also find a generalization of this solution, 
a rotating charged black hole, see \cite{Clement},
but we consider only static black holes, for simplicity.

According to the Bekenstein-Hawking formula,
the entropy of black hole (\ref{4.11})--(\ref{4.15}) 
is 
\begin{equation}\label{4.16} 
S^{BH}={1 \over 4G}{\cal A}={\pi r_+ \over 2G}~~~, 
\end{equation} 
where $\cal A$ is the length of the black hole horizon.  
The corresponding Hawking temperature is determined by the surface 
gravity $\kappa$ of the horizon 
\begin{equation}\label{4.18}
T_H={\kappa \over 2\pi}=
\left.{(N^2)' \over 4\pi}\right|_{r=r_+}=
{r_+ \over 2\pi a^2}\left(1 -
{Gq^2a^2 \over \pi r_+^2}\right)~~~.
\end{equation}
The black hole becomes extremal ($T_H=0$) at
\begin{equation}\label{4.19}
r_+^2={G q^2a^2 \over \pi}~~~.
\end{equation}

By using definitions (\ref{4.16}) and (\ref{4.18})
one finds a variational formula
\begin{equation}\label{4.20}
\delta {\cal E}_H=T_H \delta S^{BH}+\Phi \delta q~~~,
\end{equation}
\begin{equation}\label{4.21}
{\cal E}_H=
{r_+^2 \over 8Ga^2}+ {q^2 \over 4\pi} \ln {R \over r_+}+C~~~,
\end{equation}
\begin{equation}\label{4.22}
\Phi=-A_t(R)~~~,
\end{equation}
where $R$ and $C$ are dimensional constants.
Note that in an arbitrary gauge, when the condition $A_t(r_+)$
is not imposed,
$\Phi$ is  
the difference between electric potentials at the black hole
horizon and at some point $r=R$. 

Formula (\ref{4.20}) has the form of the first law of 
thermodynamics of a Reissner-Nordstr\"om black
hole,
the parameter ${\cal E}_H$ being
identified with the energy.
If an AdS black hole has no charge one can define its energy
by using Abbott-Deser \cite{AD} generalization of the ADM mass.
This energy can be equivalently written as the integral \cite{HH}
\begin{equation}\label{4.21a}
M_{\mbox\tiny{ADM}}=-{1 \over 8\pi G}\int_{C_R} N(^1K-^1\bar{K})~~~
\end{equation}
computed at the spatial circular boundary $C_R$ at $r=R$.
Here 
$^1K=N(R)/R$ is the extrinsic curvature of
$C_R$. Definition (\ref{4.21a}) requires 
a subtraction of a
"reference background" mass. This results in the term
in (\ref{4.21a}) depending on the "reference" extrinsic  curvature
$^1\bar{K}$ . In the considered case the natural choice
for the "reference background" is
the anti-de Sitter space which is another solution of AdSM-gravity
\begin{equation}\label{4.23}
d\bar{s}^2=-{N^2(R) \over \bar{N}^2(R)}
N_0^2(r)dt^2+\bar{N}^{-2}dr^2+r^2d\varphi^2~~~,
\end{equation}
\begin{equation}\label{4.24}
\bar{N}^2(r)=1+{r^2 \over a^2}~~~.
\end{equation}
The normalization of the Killing vector $\partial_t$ 
coincides at $r=R$ with normalization of $\partial_t$ for the
black hole space-time.  Now, if one adopts (\ref{4.21a}) 
also as the definition of the energy of a charged black hole
and chooses
$C=-1/(8G)$ in (\ref{4.21}),
\begin{equation}\label{4.25}
{\cal E}_H=M_{\mbox\tiny{ADM}}~~~.
\end{equation}
Two remarks are in order regarding this equation.
Firstly, in three dimensions the electric potential diverges 
logarithmically at infinity. That is why the mass of the black hole 
cannot be made finite even after subtracting the "reference" AdS 
mass.
Secondly, for asymptotically anti-de Sitter spaces
the temperature redshifts to zero at infinity. Thus, the standard
interpretation of ${\cal E}_H$ 
as an internal thermodynamic energy
of a black hole is note quite correct. For the conventional formulation
of black hole thermodynamic \cite{York}  one should use the 
quasilocal energy \cite{BCM}. However, the only difference between this 
energy and ${\cal E}_H$ 
is in the redshift factor $N(R)$.

It is worth also pointing out another possible definition of the
black hole energy which will be important for us later. 
According to Wald and Iyer \cite{WI}, one can 
generally define the black hole energy in terms of the Noether charge 
and a boundary function,
\begin{equation}\label{wi1} 
M=\int_{C_R} 
\left(Q_{\mu\nu}(t)n^{\mu}u^{\nu}+NB\right)~~~.  
\end{equation} 
Here 
$Q_{\mu\nu}$ is the Noether potential associated to the Killing field 
$\partial_t$, $u^\mu$ and $n^\mu$ are unit normals to $C_R$, 
inward-pointing and future-directed, respectively. Function $B$ comes 
out from the boundary term in the action and is introduced to have a 
well-defined variational procedure. 
By using formulas of \cite{WI} one finds
for the Einstein-Maxwell theory 
(\ref{4.1}) 
\begin{equation}\label{wi2}
Q_{\mu\nu}(t)={1 \over 8\pi G} ~t_{\mu;\nu}-F_{\mu\nu}t^\lambda
A_\lambda ~~~,
\end{equation} 
\begin{equation}\label{wi3}
B=-{1 \over 8\pi G} K~~~,
\end{equation}
where $K$ is the extrinsic curvature of the spatial boundary 
of the black hole space-time (at $r=R$).
One can define the corresponding energy $\bar{M}$
for reference AdS space-time (\ref{4.23}).
For these definitions the black hole energy takes the form
\begin{equation}\label{wi4}
M'=
M-
\bar{M}=
M_{\mbox\tiny{ADM}}-q\Phi~~~,
\end{equation}
where the last term comes out from the contribution
of the gauge field in the Noether potential (\ref{wi2}).
One can also show that 
energy (\ref{wi4}) coincides with Hawking-Horowitz
\cite{HH} definition of the mass as the surface term in the Hamiltonian.
According to the general result of Wald and Iyer \cite{WI},
the first law of black hole thermodynamics looks as follows
\begin{equation}\label{wi5}
\delta M=T_H\delta S^{BH}~~~.
\end{equation}
Actually, there is no contradiction between this formula
and the first law (\ref{4.20}). Equation (\ref{wi5}) holds
for boundary conditions which require fixing
the value of the potential $\Phi$ on the boundary.
By taking into account (\ref{wi4}) and the fact that 
$\bar{M}$ is fixed one can obtain (\ref{4.20}) from (\ref{wi5}).

\section{Charged fields near a charged black hole}
\setcounter{equation}0

We begin with the discussion of some features related to interaction of 
a charged field and electric field of a black hole
near the horizon.  
For simplicity we will be
dealing again with static black holes only. However, our basic 
conclusions will hold for rotating black holes as well. 
We also restrict the analysis by non-extremal black holes,
and then comment on extremal ones.
The metric and 
the gauge potential can be taken in the form 
\begin{equation}\label{1.6} 
ds^2=-N^2(x)dt^2+
g_{ik}(x)dx^idx^k~~~,
\end{equation}
\begin{equation}\label{5.1}
A=A_t(x)dt~~~.
\end{equation}
The Killing horizon is the surface where  $N^2(x)=0$ and
it is assumed that
on this surface $A_t=0$. We consider a space-time with an arbitrary
dimension.
Let us investigate the spectrum of single-particle excitations of a 
scalar field described by the Klein-Gordon equation 
\begin{equation}\label{1.5}
(-D^\mu D_\mu+m^2)\phi=0~~~,~~~D_\mu=\nabla_\mu+ie A_\mu~~~,
\end{equation}
where $e$ is the charge of the field. By substituting
the wave-function with the energy $\omega$
\begin{equation}
\label{1.2}
\phi_{\omega}(t,x)=e^{-i\omega t}
\phi_{\omega}(x)~~~,
\end{equation}
one obtains the equation
\begin{equation}\label{1.13a}
H^2(\omega)\phi_{\omega}(x)=\omega^2\phi_{\omega}(x)~~~,
\end{equation}
\begin{equation}\label{1.13}
H^2(\omega)=N^2\left[
\Delta_x+
V(\omega)\right]~~~,
\end{equation}
\begin{equation}\label{1.8}
\Delta_x\equiv-{1 \over \sqrt{-g}}\partial_i
\sqrt{-g}g^{ik}\partial_k~~~,
\end{equation}
\begin{equation}\label{1.13b}
V(\omega)=m^2-2\omega e A_t N^{-2}-e^2A_t^2N^{-2}~~~.
\end{equation}
Equation (\ref{1.13a})
has the form of the relativistic Schroedinger
equation with a specific "Hamiltonian" $H(\omega)$ 
where potential term (\ref{1.13b}) depends on the
frequence $\omega$.  
For antiparticles one obtains the same 
potential term (\ref{1.13b})
where charge $e$ should be replaced by $-e$ (or $\omega$ by $-\omega$).

In general, solution of problem (\ref{1.13a}) may be quite 
complicated. However, the effect of the gauge interaction
near the horizon is easy to understand. 
In the vicinity of the
horizon of a nonextremal black hole
\begin{equation}\label{5.3} 
N^2\simeq 
\kappa^2\rho^2~~~,
\end{equation} 
\begin{equation}\label{5.4} 
A_t\simeq-{\kappa \over 2} E_+\rho^2~~~,
\end{equation}
\begin{equation}\label{5.5}
V(\omega)\simeq m^2(\omega)=m^2+e\omega E_+\kappa^{-1}~~~,
\end{equation}
where $\kappa$ is the 
surface gravity constant and $\rho$ is the proper distance
to the horizon (located at $\rho=0$).   

\bigskip
\noindent
Therefore, {\it interaction near the horizon
with the electric field of the black hole
shifts effectively masses of fields}.

\bigskip

\noindent
The parameter $E_+$ in (\ref{5.4}) is the strength of the electric
field
on the horizon,
\begin{equation}\label{5.6}
E_+=l^\mu p^\nu F_{\mu\nu}~~,
\end{equation}
where $l^\mu$ and $p^\mu$ are two mutually orthogonal normals
to the bifurcation surface, $l^2=-p^2=1$,
$p^\mu$ is future directed.
For charged black hole (\ref{4.14}), (\ref{4.15})
\begin{equation}\label{5.6a}
E_+={q \over 4\pi r_+}~~~.
\end{equation}
Now, we can take into account that
quantum fields which are in a thermal equilibrium with the black
hole have to be at the Hawking temperature $T_H=\kappa/2\pi$. Thus,
the main contribution into observable quantities comes
from frequences $\omega\simeq T_H$  while
contribution from 
$\omega\gg T_H$ 
is exponentially small. 
It means that in order to estimate the effect one 
can assume that
\begin{equation}\label{5.8}
H^2(\omega)\simeq\kappa^2 \rho^2\left[
\Delta_x
+\tilde{m}^2\right]~~~,
\end{equation}
\begin{equation}\label{5.9}
\tilde{m}^2=m^2+\sigma eE_+~~~,
\end{equation}
where $\sigma\simeq O(1)$ is
a numerical coefficient. Correspondingly,
for antiparticles the
charge $e$ in $\tilde{m}^2$ is replaced by $-e$.

The effective mass $\tilde{m}$ depends on the strength
of the electric field. If the electric field
is strong, $|E_+|\gg m^2/|e|$, it
creates particle--anti-particle pairs and this process
results in the instability
of the quantum state. 

In what follows we assume that the field is 
weak and there is no pair creation process. As we will see, this
condition is satisfied in the induced gravity in the
"low energy" limit.
In this case,
the physical picture is basically the same as for neutral
black holes or uncharged fields
\cite{FF:98} and the shift of the masses is 
an irrelevant effect.
In particular by using results for uncharged fields
one concludes that the
spectrum of energies $\omega$ is continuous and does not have the
mass gap. The density of the energy levels $dn /d\omega$,
of $H(\omega)$
is divergent near the horizon, and this
results in the divergence of the entropy
of quantum fields. The leading divergence, however, is the same
as for uncharged fields.

Now we briefly comment on extremal black holes. In this case
the behavior of charged fields is different.
Near the horizon of an extremal black hole
$N^2\sim (r-r_+)^2$ and $A_t \sim (r-r_+)$,
see, e.g., Eqs. (\ref{4.14}), (\ref{4.15}) and (\ref{4.19}). Thus,
the ratio $A_t/N^2$ is singular at $r=r_+$, and the effect
cannot be described by
shift of the mass (\ref{5.5}).
Instead, behavior of a charged particle near the horizon is similar
to moving in a flat space with the   
potential term $\pm |q\omega A_t|$. However, because this potential
vanishes it cannot seriously change the 
spectrum of single-particle Hamiltonian (\ref{1.13}) and
bring new features into the considered problem.
Anyway, the Hawking temperature of extremal black holes is zero
and it seems
there is no much sense in considering statistical-mechanical entropy 
of the constituents.
We will not be discussing extremal black holes anymore.

\section{Black hole entropy in AdSM gravity}
\setcounter{equation}0

We are now ready to discuss statistical mechanical interpretation
of the black hole entropy $S^{BH}$ in induced Einstein-Maxwell gravity.
We relate $S^{BH}$ to the entanglement entropy $S^{SM}$
of the constituent fields. To regularize the
divergences caused by the horizon we use the Pauli-Villars
regularization and introduce for the each constituent with the
mass $m$ three additional fields, one with the normal statistics
and the  mass $M_1=\sqrt{m^2+2\mu^2}$ and two with the wrong statistics
and the masses  $M_2=\sqrt{m^2+\mu^2}$. The parameter $\mu$ is the
Pauli-Villars cutoff.

The analysis of the divergences in 
three dimensions is similar to other dimensions,
see, e.g., \cite{F:98},\cite{FF:98}. It can be shown that
in three dimensions
only the leading divergences are present. By following the
method of \cite{F:98} one immediately finds the
regularized values of the densities of the energy levels
\begin{equation}\label{6.1}
\left({dn \over d\omega}\right) ={\tilde{b}(\mu,m) \over 2\pi \kappa} 
{\cal A} ~~~,
\end{equation} 
for a charged scalar or spinor constituent. 
Here ${\cal A}=2\pi r_+$ is the length of the black hole horizon. 
Expressions (\ref{6.1}) include for each field contributions
of particles and antiparticles.  
At large $\mu$ 
\begin{equation}\label{6.3}
\tilde{b}(\mu,m)=2\gamma\mu 
-m(\omega)-m(-\omega)~~,~~\gamma=2-\sqrt{2}~~~,  
\end{equation} 
where $m(\omega)$ and $m(-\omega)$ are
effective masses of particles and antiparticles
determined in (\ref{5.5}).

In the "low energy" limit of induced gravity
the effect of shifting the masses  
is negligibly small.
According to (\ref{A.14}), the charges of the constituents $e_k$
should be restricted from above,
$|e_k| <\sqrt{m_k}$. Thus, 
as follows from (\ref{4.7}),
\begin{equation}\label{6.4} 
|e_k E_+|\ll m_k^2~~~,
\end{equation} 
which guarantees that there is no pair creation 
by the electric field of the black hole.  By using this one can also
rewrite (\ref{6.3}) at $\omega < T_H$ as
\begin{equation}\label{6.3a}
\tilde{b}(\mu,m)\simeq 2\gamma\mu-2m+O(\omega^2)~~~.
\end{equation} 
At low energies the terms quadratic in $\omega$ and higher 
result in small corrections and we neglect them in computations.

The entropy can be found from the free 
energy, $S=\beta^2\partial_\beta F[\beta]$, where
\begin{equation}\label{1.24}
F[\beta] = \eta\beta^{-1}\int d\omega {dn(\omega) \over d\omega}
\ln (1-\eta e^{-\beta \omega})~~~,
\end{equation}
$\eta=+1$ for bosons and $-1$ for fermions.
By using Eqs. (\ref{1.24}), (\ref{6.1}) and 
(\ref{6.3a}) one finds the entropy
\begin{equation}\label{6.5}
S_s \simeq {1 \over 6}(\gamma\mu -m_s) {\cal A}~~~,~~~
S_d \simeq {1 \over 12}(\gamma\mu -m_d) {\cal A}~~~,
\end{equation}
for scalars and spinors, respectively.
The total entropy of the constituents is
\begin{equation}\label{6.6}
S^{SM}=\sum_s S_s+\sum_d S_d=
{1 \over 12}
\left(\gamma \mu(2N_s+N_d)-2\sum_s m_s-\sum_d m_d\right){\cal A}~~~.
\end{equation}

According to the induced gravity relation (\ref{i.1}), the non-minimally
coupled constituents give  additional contribution to the
Bekenstein-Hawking entropy in the form of the Noether charge 
\cite{FFZ}
\begin{equation}\label{6.7}
Q=2\pi \sum_s \xi_s \int_{\Sigma}\langle (\phi_s)^+\phi_s \rangle~~~,
\end{equation}
where the field operators are taken on the horizon.
The charge $Q$ is ultraviolet-divergent and in the
Pauli-Villars regularization
\begin{equation}\label{6.8}
Q= \left(\gamma \mu \sum_s \xi_s-\sum_s m_s\xi_s\right)
{\cal A}~~~.
\end{equation}
Therefore,
\begin{equation}\label{6.9}
S^{SM}-Q={1 \over 12}\left(\gamma \mu(2N_s+N_d-12\sum_s\xi_s)
+2\sum_s(6\xi_s-1)m_s-\sum_d m_d\right){\cal A}~~~.
\end{equation}
If the induced gravity constraints (\ref{4.6}) are satisfied,
the divergence of the Noether charge $Q$ compensates the divergence
of the entropy $S^{SM}$ and the following
identity
\begin{equation}\label{6.10}
S^{SM}-Q={1 \over 4G}{\cal A}=S^{BH}~~~,
\end{equation}
where $G$ is induced gravitational constant (\ref{4.9}), takes place.
Therefore, the induced gravity relation (\ref{i.1})
holds for charged black holes as well.

\bigskip

The subtraction in entropy formula (\ref{6.10})
has the same interpretation as for 
neutral black holes in four-dimensional 
induced gravity \cite{FF:97}, \cite{FF:99a}.
To see this,  
it should be noted first that the Bekenstein-Hawking entropy of a 
charged  black hole in induced AdSM
gravity is related to 
the spectrum of the black hole mass $M$ defined by
Eq. (\ref{wi1}).
Indeed, consider a 
small excitation of constituent fields with energy 
${\cal E}$
over a vacuum (${\cal E}=0$). 
Such an excitation 
results in a change of black hole parameters.
The corresponding variational formula was studied in
\cite{F:98b} \footnote{It is worth  mentioning
that Eq. (\ref{3.55})
was derived in Ref. \cite{F:98b} for uncharged black holes.
However, if the mass of the black hole is defined by (\ref{wi1}),
formula (\ref{3.55}) takes place for charged black holes as well.
Equation (\ref{3.55}) holds in the linear order in perturbations 
provided that the background
black hole metric and the gauge potential satisfy the 
equations of motion. Also, the
variations of matter fields have to vanish
on the boundary of the black-hole space-time. This requirement,
however, is not important because the constituent fields
are trapped inside the potential barrier and can be excited only in
a thin layer near the horizon. }
\begin{equation}\label{3.55}
\delta M=T_H\delta S^{BH}
+{\cal E}~~~,
\end{equation}
where the energy  
\begin{equation}\label{3.66}
{\cal E}=\int_{\Sigma_t} T^{\mu\nu}t_\mu d\Sigma_\nu~~~,
\end{equation}
is determined
in terms of the stress-energy tensor of the constituents
$T^{\mu\nu}$ 
($\Sigma_t$ is the hypersurface of constant time $t$,
$d \Sigma_t$ is the future-directed vector of the volume element
of $\Sigma_t$, and $t_\mu$ are the components of 
$\partial_t$).

Thus, for a black hole with the fixed area the spectrum of $M$ 
is related to the spectrum of energies ${\cal E}$
of the constituents. 
On the other hand, the 
statistical-mechanical entropy $S^{SM}$ of the
constituents is determined by the 
spectrum of their Hamiltonian ${\cal H}$ which
generates canonical transformations 
of the system along the Killing field $\partial_t$. 
The observation crucial for
understanding entropy relation (\ref{6.10}) is
that the energy ${\cal E}$ and the Hamiltonian ${\cal H}$ 
of the
non-minimally coupled constituents differ by a total derivative
which picks up a non-vanishing contribution on the 
inner boundary $\Sigma$ of $\Sigma_t$, i.e., on the horizon.
The boundary term is 
the Noether charge on $\Sigma$
\begin{equation}\label{3.67}
{\cal H}-{\cal E}=T_H Q~~~,
\end{equation}
where $T_H$ is the Hawking temperature.
It is because of Eqs. (\ref{3.55}), (\ref{3.66}), (\ref{3.67}) we 
expect that the two entropies, $S^{BH}$ and $S^{SM}$, are different and 
related by (\ref{6.10}).

\bigskip

Now a remark concerning charged rotating black holes
is in order. This sort of black hole solutions in AdSM gravity
was found in Ref. \cite{Clement} and represents a generalization
of static solution (\ref{4.11})--(\ref{4.15}). 
In the corresponding induced gravity
the constituent fields co-rotate together with a black 
hole\footnote{Because the constituents are
very heavy and trapped near the horizon, they 
are automatically inside the 'null' cylinder.}. In the
corotating frame of reference one can define a canonical ensemble
of the constituents and compute their entropy $S^{SM}$.
Obviously, in the corotating frame the properties
of the constituents are as if the black hole were static and,
in the leading order, $S^{SM}$ is given by (\ref{6.6}).
(One can verify this by doing a more rigorous analysis, 
see Ref. \cite{FF:99a}.)
Therefore, in induced AdSM gravity
the entropy of rotating charged black holes is 
still expressed by formula (\ref{6.10}).
Also, the above interpretation of (\ref{6.10}) 
can be extended to take into account the rotation, see again
Ref. \cite{FF:99a}.

\section{Concluding remarks}

In conclusion we comment on
black holes in four-dimensional induced Einstein-Maxwell 
gravity\footnote{Constructing
such a theory with charged constituents
would 
require cancellation of additional ultraviolet divergences,
and it would be more complicated than constructing
AdSM gravity in three dimensions or induced pure Einstein gravity 
in four dimensions.}. 
There is a number of reasons to believe that
basic features of the entropy of charged black holes in 
four dimensions will be similar to properties of the entropy
of charged black holes in 
AdSM gravity.

To see what is happening in four dimensions let us consider
first the 
density of energy levels of an uncharged scalar
field computed, say, in Pauli-Villars regularization \cite{FF:98}
\begin{equation}\label{7.1} 
\left[{dn (\omega|\mu)\over d\omega}\right]
={1 \over (4\pi)^2\kappa}\int_{\Sigma}\left[2b+a
\left({\omega^2 \over \kappa^2}{\cal P}+2\left(\frac 16-\xi\right) R
\right)\right]~~~.
\end{equation}
Here the integral
is taken over the bifurcation surface $\Sigma$ of the horizon.
The quantity $R$ is the scalar curvature of the black hole
space-time computed on $\Sigma$, 
\begin{equation}\label{7.2}
{\cal P}=2{\cal R}-{\cal Q}~~~,~~~{\cal Q}=P^{\mu\nu}R_{\mu\nu}~~~,
~~~{\cal R}=P^{\mu\nu}P^{\lambda\rho}R_{\mu\lambda\nu\rho}~~~,
\end{equation}
where $P^{\mu\nu}=l^\mu l^\nu-p^\mu p^\nu$ is a projector onto a two-dimensional
surface orthogonal to $\Sigma$, and $p^\mu$, $l^\mu$ are
two mutually orthogonal normals of $\Sigma$ ($l^2=-p^2=1$).
The regularization parameter $\mu$ defines the scale of the Pauli-Villars
masses, and at large $\mu$
\begin{equation}\label{7.3}
a\simeq \ln{\mu^2 \over m^2}~~~,~~~
b\simeq \mu^2\ln {729 \over 256} -m^2\ln {\mu^2 \over m^2}
\end{equation}
where $m$ is the mass of the field (see for details \cite{FF:98}).

As we saw in Section 4, if the field is charged, its interaction 
with the electric field of the black hole results in the shift of the
mass determined by Eq. (\ref{5.5}). 
The total density of energy levels of particles and
antiparticles will be given by (\ref{7.1})
where, according with (\ref{5.5}) and (\ref{7.3}), one has 
to replace constant $a$ and $b$  by 
\begin{equation}\label{7.4}
a(\omega)=\ln{\mu^2 \over m^2(\omega)}+\ln{\mu^2 \over 
m^2(-\omega)}~~~, 
\end{equation}
\begin{equation}\label{7.5}
b(\omega)=
2\mu^2\ln {729 \over 256} -
m^2(\omega)\ln {\mu^2 \over m^2(\omega)}-
m^2(-\omega)\ln {\mu^2 \over m^2(-\omega)}~~~.
\end{equation}
These expressions are even functions of $\omega$ and,
as a consequence, they do not depend on $\omega$ at large $\mu$.
The similar changes in the density 
of levels take place for charged spinor fields.

As follows from (\ref{7.4}), (\ref{7.5}), the interaction of 
charged particles with the electric field does not change the 
ultraviolet divergence of the density of levels near the horizon.
The electric field strength $E_+$ appears only in finite corrections
to  $dn /d\omega$.
Analogously, {\it the divergences of the statistical entropy}
$S^{SM}$ {\it of fields 
does not depend on the gauge interaction and remain purely
geometrical}.  This property is 
the same as in case of the three-dimensional charged black holes.

What properties may one
expect in induced Einstein-Maxwell gravity in four dimensions?
The Bekenstien-Hawking entropy $S^{BH}$ of a charged black hole is one 
quarter of the horizon area $\cal A$ regardless the presence of 
gauge fields. Also the entropy $S^{SM}$ of the constituents 
and the corresponding Noether charge 
remain proportional to $\cal A$ in the leading order
and do not depend on $E_+$. That is why one can conclude that
the entropy  $S^{BH}$ for charged and neutral black holes has the 
universal form (\ref{i.1}).

\bigskip

This sort of universality is very important and, as we stressed earlier 
\cite{FFZ}, it demonstrates that the mechanism of generation of the 
Bekenstein-Hawking entropy is a low-energy phenomenon which
depends neither  on the properties of a black hole nor on the
specific structure of an underlying fundamental theory of gravity
(on the number of species of constituents and their parameters, for 
example).  

\bigskip

Our analysis holds for near extremal black holes
whose thermodynamical behavior is known to be
similar to properties of a two-dimensional massless quantum gas.
It is intriguing problem to understand on the level of the
induced gravity constituents how this effective two-dimensional 
description becomes possible. 
Another aspect of near-extremal black holes is that
they can be described by an "effective string theory" 
(by a 2D supersymmetric conformal field theory) see, e.g., 
\cite{MS}.  Thus, at this point the induced gravity and
string theory derivations of the black hole entropy
overlap and one has a chance to explore whether there is
any correspondence between the two pictures.

\vspace{12pt}
{\bf Acknowledgements}:\ \ 
The work of V.F. is 
partially supported by the Natural
Sciences and Engineering
Research Council of Canada and by the Killam trust.
D.F. is supported in part by the RFBR grant
N 99-02-18146. This work is done in framework of NATO collaboration.

\newpage

\appendix
\section{Appendix}
\setcounter{equation}0

Here we comment on how Eqs. (\ref{4.6}), (\ref{4.9}), (\ref{4.10})
can be obtained. For simplicity we consider
Euclidean theory and use the standard 
representation 
\begin{equation}\label{A.1} 
W=-{\eta \over 2}\int_{\delta}^{\infty}{ds \over s} \mbox{Tr}(e^{-s L})
e^{-m^2s}
\end{equation}
for the regularized
one-loop effective action
\begin{equation}\label{A.2}
W=\eta\frac 12 \log \det(L+m^2)
\end{equation}
with $\eta=1$ and $\eta=-1$ for boson
and fermion fields, respectively.
Here $L$ is the wave operator, $L_s=-D^\mu D_\mu +\xi R$
for scalar constituents, and $L_d=-D^\mu D_\mu +\frac 14 R$
for spinor ones. The covariant derivative 
is $D_\mu=\nabla_\mu +e A_\mu$.
The parameter $\delta$ is the
ultraviolet cutoff.

When the mass $m$ of the field is sufficiently
high it is enough to approximate $W$ by the local expansion over the
mass parameter.
To this aim one replaces the trace of the heat kernel of $L$
in (\ref{A.1}) by the asymptotic expansion over $s$ with
the heat kernel
coefficients $a_n$. In three dimensions the calculation
gives
\begin{equation}\label{A.3}
W_\delta=-
{\eta \over 16 \pi^{3/2}}\left[m^3 a_0\Gamma(-3/2,m^2\delta)+
ma_1\Gamma(-1/2,m^2\delta)+\pi^{1/2}{1 \over m} a_2+O(m^{-3})\right]~~~,
\end{equation}
\begin{equation}\label{A.4}
\Gamma(z,x)=\int_{x}^{\infty}dt t^{z-1} e^{-t}~~~.
\end{equation}
With the following asymptotics for the incomplete gamma-function
\begin{equation}\label{A.5}
\Gamma(-1/2,x)\simeq 2x^{-1/2}-2 \pi^{1/2}~~~,
\end{equation}
\begin{equation}\label{A.6}
\Gamma(-3/2,x)\simeq\frac 23 x^{-3/2}
-2 x^{-1/2}+{4 \over 3} \pi^{1/2}~~~,
\end{equation}
one gets
\begin{equation}\label{A.7}
W_\delta=W_\delta^{\mbox{\tiny{div}}}+
W^{\mbox{\tiny{reg}}}~~~,
\end{equation}
\begin{equation}\label{A.8}
W_\delta^{\mbox{\tiny{div}}}=-{\eta \over 16\pi^{3/2}}
\left[\frac 23 \delta^{-3/2}a_0+2 \delta^{-1/2}(a_1-m^2 a_0)
\right]~~~,
\end{equation}
\begin{equation}\label{A.9}
W^{\mbox{\tiny{reg}}}
=\eta\left[-{1 \over 12\pi} m^3 a_0+
{1 \over 8\pi} ma_1-{1 \over 16\pi}{1 \over m} a_2 +O(m^{-3})
\right]~~~.
\end{equation}
The heat coefficients $a_0$ and $a_1$ do not depend
on the gauge field $A_\mu$. For charged scalars
\begin{equation}\label{A.10}
a_0=2\int\sqrt{g}d^3x~~~,~~~a_1=2(\frac 16-\xi)\int\sqrt{g}d^3x
R~~~,
\end{equation}
and for three-dimensional spinors
\begin{equation}\label{A.11}
a_0=2\int\sqrt{g}d^3x~~~,~~~a_1=-\frac 16\int\sqrt{g}d^3x
R~~~,
\end{equation}
The next heat-kernel coefficient has the following form
\begin{equation}\label{A.15}
a_2=\int \sqrt{g}d^3x\left[{1 \over 180}(c_1 R^{\mu\nu}R_{\mu\nu}
+c_2 R^2)+c_3 F^{\mu\nu} F_{\mu\nu}\right]~~~.
\end{equation}
In this equation we neglect total derivatives and took into account that in three dimensions
the Riemann tensor is
\begin{equation}\label{A.16}
R_{\mu\nu\lambda\rho}=g_{\mu\rho} R_{\nu\sigma}+
g_{\nu\sigma} R_{\mu\rho}-g_{\nu\rho}R_{\mu\sigma}-
g_{\mu\sigma} R_{\nu\rho}
-\frac 12 (g_{\mu\rho} g_{\nu\sigma}-g_{\mu\sigma} g_{\nu\rho}) R
~~~.
\end{equation}
The coefficients are: $c_1=6$, $c_2=6+5(1-6\xi)^2$,
$c_3=-e_s^2/6$ for a scalar field with the charge $e_s$ and non-minimal
coupling $\xi$, see, e.g., \cite{DeWitt},
and $c_1=-9$, $c_2=-4$ , $c_3=e_d^2/3$ for a 3D spinor field
with the charge $e_d$.

The induced gravity requires cancelation of the divergences
\begin{equation}\label{A.13}
\sum_sW_{\delta,s}^{\mbox{\tiny{div}}}
+\sum_dW_{\delta,d}^{\mbox{\tiny{div}}}=0~~~.
\end{equation}
Equation (\ref{A.13}) is equivalent to three conditions
(\ref{4.6}) which can be satisfied for a number of models.
The total induced action is determined then by the
contributions of the regular parts $W^{\mbox{\tiny{reg}}}$,
Eq. (\ref{A.9}), only.
Let us now impose the additional restriction on the
charges of the constituents
\begin{equation}\label{A.14}
\sum_s {e_s^2 \over m_s}  +
2\sum_d{e_d^2 \over m_d}=24\pi~~~
\end{equation}
which provides the right coefficient by the Maxwell action.
Then the induced Euclidean action in the leading order in curvature and
the strength of the gauge field has the form
\begin{equation}\label{A.17}
\Gamma[g,A]=-\frac 14 \int d^3x\sqrt{g}\left[{1 \over 4\pi 
G}(R-2\Lambda) -F^{\mu\nu}F_{\mu\nu}+aR^{\mu\nu}R_{\mu\nu}+b 
R^2\right]~~~.  
\end{equation}
with
the induced gravitational and cosmological constants defined
by Eqs. (\ref{4.9}), (\ref{4.10}) and
\begin{equation}\label{A.18}
a={1 \over 240\pi}\left(2\sum_s {1 \over m_s}+3\sum_d {1 \over m_d}
\right)~~~,
\end{equation}
\begin{equation}\label{A.19}
b={1 \over 720\pi}\left(\sum_s \left(6+5(1-6\xi_s)^2\right)
{1 \over m_s}+4\sum_d {1 \over m_d}
\right)~~~.
\end{equation}
Functional (\ref{A.17}) contains higher curvature terms  which
can be neglected in low-energy limit. Note that
in this limit
the terms  quadratic in curvature can be also neglected as compared
with the "Einstein-Maxwell" part. Then 
after the Wick rotation functional (\ref{A.17}) coincides
with the Lorentzian action (\ref{4.1}).

\end{document}